\documentstyle[twocolumn,aps,prl]{revtex}
\input epsf.sty
\begin{document}
\draft
\preprint{}
\title{Anharmonic effects in the A15 compounds induced by sublattice distortions}
\vskip -1.6cm
\author{Z. W. Lu and Barry M. Klein}
\address{Department of Physics, University of California, 
Davis, California 95616}
\date{To appear in Phys. Rev. Lett.}
\address{\mbox{ }}
\address{\parbox{14cm}{\rm \mbox{ }\mbox{ }
We demonstrate that elastic anomalies and lattice instabilities in the 
the A15 compounds are describable in terms of first-principles LDA electronic 
structure calculations.  We show that at $T=0$ V$_3$Si, V$_3$Ge, and Nb$_3$Sn 
are intrinsically unstable against shears with elastic moduli 
$C_{11}-C_{12}$ and $C_{44}$, and that the zone center phonons, 
$\Gamma_2$ and $\Gamma_{12}$, are either unstable or extremely soft.  
We demonstrate that sublattice relaxation (internal strain) effects 
are key to understanding the behavior of the A15 materials.
}}
\address{\mbox{}}
\address{\parbox{14cm}{\rm PACS numbers: 62.20Dc, 74.25.Jb, 74.25.Kc, and 74.70Ad}}
\maketitle
\newpage

\narrowtext

The A15 materials have been of interest for more than 40 years due to the 
relatively high superconducting transition temperatures for many of them and 
the interplay between their superconductivity, their structural phase 
transitions, and their temperature-dependent anomalies in properties such as 
elastic constants, phonon spectra, electrical resistivity, Knight shift and 
magnetic susceptibility.\cite{test73,wege73,izyu74}
The paradigm interesting A15 materials are 
V$_3$Si and Nb$_3$Sn, both of which have relatively high $T_{\rm c}$ values 
and also undergo cubic-to-tetragonal structural phase transitions just above 
$T_{\rm c}$ (with $\frac{c}{a} > 1$ and $\frac{c}{a} < 1$, respectively), and 
have many other anomalous properties as well.  Many theoretical models, 
discussed in Refs. [1-3], attempt to explain the unusual properties of the A15
materials in terms of models with a sharply peaked structure in the electronic
density of states (DOS) near the Fermi level, $E_{\rm F}$.

Although the past ten years has seen a major focus on developing an 
understanding of the ceramic oxide superconductors,\cite{oxide} there remain many 
unanswered fundamental questions related to the unusual properties of the A15
materials, including the question of the relationship between their high 
$T_{\rm c}$ and their structural anomalies/instabilities, and whether or not 
they are driven 
directly by the electronic distribution using a conventional Fermi liquid 
description of the electronic structure, or whether ``exotic'' excitations are 
needed to explain the properties.  Some of these issues may have 
relevance to the high $T_{\rm c}$ oxides as well, where structural 
instabilities are also known to exist.\cite{oxide}  

Here we show that using state-of-the-art first-principles electronic structure
methods to study three of the most interesting A15 materials: V$_3$Si,
Nb$_3$Sn, and V$_3$Ge.  We have found several provocative results, 
including the facts that: (1) for all three materials the calculated elastic 
constants agree very well with the observed room temperature values if 
sublattice relaxation effects are frozen out; (2) when sub-lattice relaxation 
effects are included, the crystals are unstable in the cubic A15 structure 
with respect to a tetragonal distortion which stabilizes both the elastic 
constants and the phonon anomalies; (3) the structural instability is weakest 
(in terms of the energetics) in V$_3$Ge and may not be observable due to 
effects of crystal imperfections; (4) anomalies in several of the 
zone-center phonon branches occur along with the elastic anomalies, the 
branches corresponding to modes where atoms in the chains vibrate in the chain
direction.

The electronic structure results reported in this Letter were generated using 
the all-electron, full-potential linearized-augmented-plane-wave (LAPW) method
in the local density approximation (LDA). The valence/conduction
band states were treated semi-relativistically (no spin-orbit effects) while 
the core states were treated fully relativistically.  Care was taken to ensure 
appropriate k-point convergence for the accuracy needed in these studies.

Usually it is found that the elastic constants, {\it when evaluated at the
experimental volume}, are within 5-10\%  
of the experimental 
values.\cite{mehl94} For a cubic system (such as the A15) to be 
mechanically stable, the bulk modulus ($B \equiv \frac{C_{11}+2C_{12}}{3}$) 
and shear elastic constants ($C_{11}-C_{12}$ and $C_{44}$) must be 
positive.  The overall phonon spectrum must be stable as well (positive square 
frequencies). The elastic constants can be extracted using the procedures 
discussed by Mehl {\it et al.}:\cite{mehl94}
(i) calculate the total energy as a function of the volume, and then fit
the results to an equation-of-state 
to extract the bulk modulus $B$; 
\begin{figure}
\hskip 1.5cm
\epsfxsize=5.0cm
\epsfbox{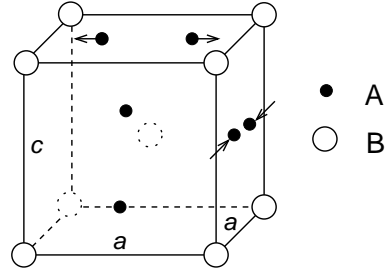}
\vskip 0.5cm
\caption{Schematic diagram of ideal ($c=a$)
and tetragonally distorted ($c \ne a$) A15 crystal structures
($A_6B_2$).  $B$ atoms form a body-centered cube, while $A$ atoms
form linear chains on each cube face.  Arrows denote allowed internal degrees of freedom
(direction of sublattice relaxations) in the tetragonally distorted A15 
structure.
}
\label{f-structure}
\end{figure}

\noindent (ii) apply an appropriate strain to the lattice to distort the primitive
lattice vectors, calculate the total energy versus strain $\delta$ at volume 
$V$, and determine the second-order elastic constants from the coefficient 
second order in the strain. For example, using an orthorhombic 
distortion, the distortion energy is,
\begin{equation}
\Delta E(\pm \delta)=V(C_{11}-C_{12})\delta^2
+O(\delta^4)\;.
\label{e:ortho}
\end{equation}
We can thus extract $C_{11}-C_{12}$ from Eq.~(\ref{e:ortho}).
Similarly, we used a monoclinic distortion to extract $C_{44}$.

Our calculated total 
energies of V$_3$Si, V$_3$Ge, and Nb$_3$Sn in the A15 structure as a function
of volume, were used to 
extract the equilibrium volumes and bulk moduli of these materials.  The 
calculated equilibrium volumes are underestimated by 4-6\%, typical of LDA 
calculations, 
leading to the calculated equilibrium bulk moduli being overestimated compared to 
experiment by more than $\sim$ 25\%. On the
other hand, the calculated bulk moduli at the {\it experimental volumes}
are very close ($\sim$ 5\%) to those of experiment, as shown in 
Table~\ref{t-elastic}.

\begin{figure}
\hskip 0.5cm
\epsfxsize=7.0cm
\epsfbox{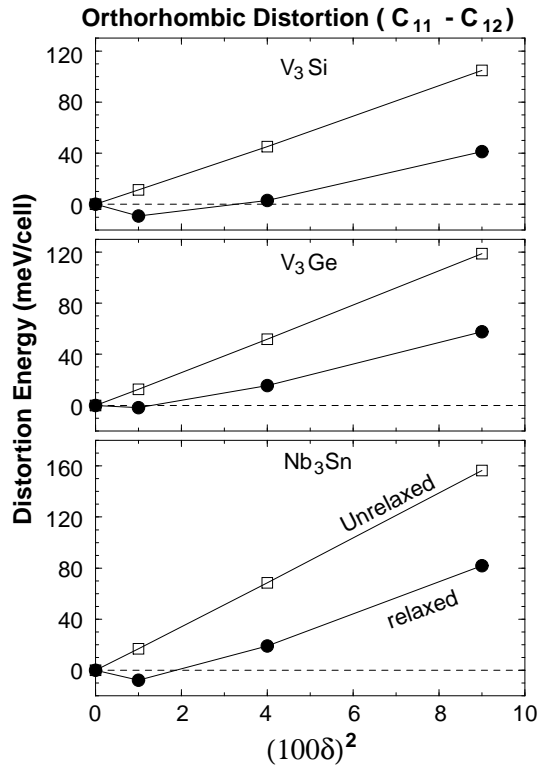}
\vskip 0.5cm
\caption{Distortion energies for orthorhombically distorted A15 structure.
Calculations were performed at the experimental A15 volumes.  The open
squares denote calculation performed with A15-like internal coordinates, while
filled circles denote calculations with internal coordinates fully relaxed.}
\label{f-ortho}
\end{figure}

Next, we examined the stability of the A15 materials under the application of
shears corresponding to the elastic constants $C_{11}-C_{12}$
and $C_{44}$, first without allowing the cell-internal sublattice 
displacements that are allowed by the broken symmetries produced by the 
elastic distortions. Figure~\ref{f-ortho} depicts our calculated distortion 
energies (open squares) as a function of $\delta^2$ using orthorhombically
distorted cells.  For a perfectly harmonic crystal, the data points
would fall on a straight line, with the slopes given by $C_{11}-C_{12}$.
Indeed the calculated points fall quite close to straight lines, indicating
only small contributions from higher order terms [$O(\delta^4)$]. The fitted
$C_{11}-C_{12}$ and $C_{44}$ values, shown in Table~\ref{t-elastic}, 
are very close to the experimental room temperature values, rather than to the low
temperature values as might be expected for a typical material.  Since first-principle calculations are supposed to 
correspond to $T=0$, not to high temperatures, why don't we reproduce 
the soft elastic constants seen at low temperature?

This puzzle was solved by realizing that as one distorts a crystal from
a higher to a lower symmetry space group, new internal degrees of 
freedom could be introduced with significant physical and computational 
importance, especially  for systems with lattice instabilities.  We note that 
it is tempting to overlook these sublattice distortions because 
determining their effect introduces a large increase in the computational 
burden, since for each internal distortion a set of self-consistent 
calculations needs to be performed to find the equilibrium sublattice 
distortions corresponding to the volume and strain under investigation.  

For the orthorhombically distorted A15 cell, the sublattice displacements 
correspond to pairs of $A$ atoms moving toward or away from each other 
(rather than fixing on their symmetry mandated A15 positions), but 

\noindent different
pairs move independently.  Thus, we need to add (iii) sublattice relaxation 
into our above procedures for calculating elastic constants. After fully 
relaxing the sublattices, we have found that the relaxed 
distortion energies for the A15 compounds (denoted by 
filled circles in Figure~\ref{f-ortho}) are lowered by a substantial
amount, which increases as the distortion gets larger. The relaxed distortion
energies show a minimum around $\delta \sim $ 1--2\% and even dip below the
energies of the ``ideal'' cubic A15 structure. Hence, these three A15 compounds
are intrinsically unstable in the cubic structure at $T=0$.  Note that the 
$C_{11}-C_{12}$ instability appears to be weakest in V$_3$Ge. Again, the 
weakest $C_{44}$ instability occurs in V$_3$Ge, while the strongest 
instability occurs for Nb$_3$Sn, as seen in experiment (Table~\ref{t-elastic}).

\begin{figure}
\hskip 0.5cm
\epsfxsize=6.0cm
\epsfbox{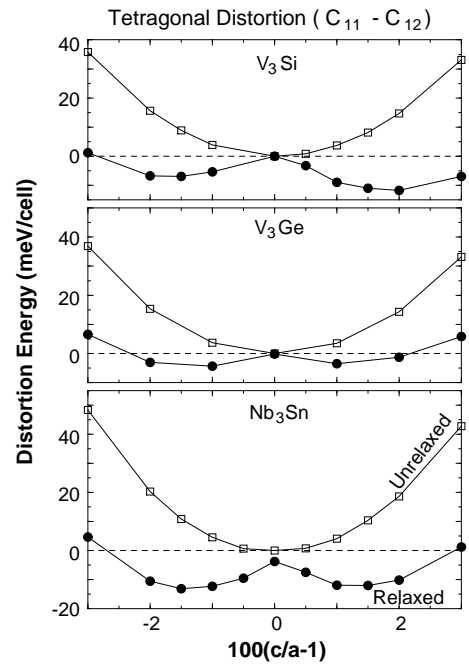}
\vskip 0.5cm
\caption{Distortion energies for tetragonally distorted A15 structure and
also see captions of Fig.~\protect\ref{f-ortho}.}
\label{f-tetra}
\end{figure}

An alternative way to obtain $C_{11}-C_{12}$ is to distort the cell
tetragonally.  This tetragonally distorted
cell turns out to have the same structure as the observed tetragonal 
phase\cite{shir71} for
some A15 compounds (Fig.~\ref{f-structure}).  The distortion energies are
depicted as a function of distortion ($\delta=\frac{c}{a}-1$) in 
Fig.~\ref{f-tetra}.
If we freeze the atoms in their A15-like coordinates, the calculated
sublattice-unrelaxed distortion energies (denoted by open squares) fall quite 
closely on a parabola.  Note that the distortion energy contains odd terms of 
distortion $\delta=\frac{c}{a}-1$ rather than being an even function 
of $\delta$ as 
in the case of orthorhombic and monoclinic distortions, so that we needed to 
perform more first-principles calculations (for $\pm \delta$) for this case. 
Indeed, we see a slight asymmetry about the $\delta=0$ axis, indicating a 
small cubic term [$O(\delta^3)$].  Least-square fitting these data to the form 
$a\delta^2+b\delta^3+c\delta^4$, we can extract $C_{11}-C_{12}=3a/V$. 
Such fitted $C_{11}-C_{12}$ values are within 6\% of those shown in 
Table~\ref{t-elastic} using orthorhombically distorted cells.

Next we allowed the internal coordinates to relax in a manner consistent with the tetragonal structure,
and found that the distortion energies were lowered by a large amount, showing
a double well structure with respect to $\frac{c}{a}$.  In the case of V$_3$Si,
the minimum occurs for $\frac{c}{a}>1$, while for Nb$_3$Sn the minimum occurs 
for $\frac{c}{a}<1$, in agreement
with experimental observations\cite{test73}. For the case of V$_3$Ge, we find an 
extremely shallow well and a much weaker instability (bordering on the accuracy 
of our LAPW calculations).  Therefore, a small amount of sample imperfection, 
such as impurities, vacancies, or disorder, or thermal effects, could inhibit the transition from 
occurring in V$_3$Ge.  Experimentally, no cubic-to-tetragonal phase transition
has been observed for V$_3$Ge, although there is some evidence for anomalies in its elastic constants \cite{rose69}.

For V$_3$Si and Nb$_3$Sn, the details of the structural phase transitions are 
very sensitive functions of sample conditions, with, for instance, the 
observed $\frac{c}{a}$ ratios, varying from 0.9964 to 0.9938 for Nb$_3$Sn
and approximately 1.0024 for V$_3$Si,\cite{test73}, while the 
calculated minimum occurs at $\frac{c}{a}$ about 1.02 and 0.985 for V$_3$Si 
and Nb$_3$Sn, respectively, significantly larger than experiment. Since the relaxed distortion energy curves
are very small and the curves are very flat, a precise determination of the $\frac{c}{a}$ 
ratio is difficult, although most of the quantitative disagreement between theory and experiment can be ascribed to LDA errors that do not affect the conclusions regarding the instabilities that we are focusing on.

We see from these results that the key to understanding the elastic constant 
softening, and most probably other anomalous properties in the A15 materials,
is the relaxation of the internal degrees of freedom (sublattice relaxation) 
when there are cell distortions, which leads to highly anharmonic behavior in 
the A15's, general ideas put forward by Testardi.\cite{test72}  At high 
temperatures, thermal motion washes away the well structure and  atoms spend 
the majority of their time away from the flat well minima and assume 
``average'' A15 positions.  Calculations with
``frozen-in'' A15-like coordinates give the maximum attainable elastic
constants, which correspond closely to the experimental room temperature
values.  As the temperature is lowered, the effect of sublattice relaxation
sets in, and at $T=0$, these A15 compounds are intrinsically unstable with
respect to $C_{11}-C_{12}$ and $C_{44}$ shear distortions.  These results can
be interpreted in terms of a ferroelastic type of instability as discussed by
Miller and Axe\cite{mill67}, with examples given in the book of 
Silje.\cite{salj90}

We have seen that the first-principles electronic structure method
predicts the existence of elastic constant instabilities and the cubic
to tetragonal phase transition in A15 materials.  It is also well established 
that the overall phonon spectra of many of the A15's exhibits unusual 
temperature dependencies and selected softening in different parts of the 
Brillouin Zone\cite{shir71,schw76,pint85}. Such 
instabilities for the phonons also follow from our calculations using the 
frozen phonon method which we have applied to determine the zone-center optic
phonons.  Note that the {\it relaxed} distortion energy in 
Fig.~\ref{f-tetra}(c) for Nb$_3$Sn at $\delta=0$ ($\frac{c}{a}=1$) 
has a small negative value of $\sim -4$ meV/cell, indicating an unstable zone 
center $\Gamma_{12}$ phonon, since this phonon has the same 
space group as the tetragonally distorted A15 cell, but with $c \equiv a$.  
The calculation for the $\Gamma_{2}$ phonon (pairs of  $A$ atoms move toward 
or away from each other with the same displacements for all three orthogonal 
chains) also shows a small, negative phonon energy for Nb$_3$Sn, indicating an 
instability.  Calculations for the $\Gamma_{12}$
and $\Gamma_{2}$ phonons for V$_3$Si and V$_3$Ge also show anomalies. However,
for these materials these phonons have negative or very small
positive energies, e.g., their phonon energies are less than 
1 meV/cell for the $\Gamma_{12}$ phonon at a displacement of
$0.003a$, where $a$ is the cubic lattice constant. For comparison, 
zone center phonon energies at the same displacement for a typical 
intermetallic compound such as Cu$_3$Pt in the cubic 
$L1_2$ structure are $\sim 25$ meV/cell (where the unit cell contains

\noindent  4 atoms). Note that
since we estimate the accuracy of our LAPW calculations for phonon energies to
be approximately 2 meV/cell, we can argue that the zone center phonons 
$\Gamma_{12}$ and $\Gamma_{2}$ are extremely soft at best, and are on the 
borderline of being stable or unstable.

First-principles electronic structure calculations have shown that the DOS
near $E_{\rm F}$ for V$_3$Si and Nb$_3$Sn have very sharp 
structures,\cite{matt65,klein78} in qualitative agreement with early model 
calculations of the DOS, and we find similar results here.  What is 
particularly interesting, however, is that we find explicitly that the DOS 
singularities, and the associated Fermi surface nesting features, are removed 
following the sublattice distortions and the resultant structural phase 
transitions. We will discuss this more fully in a future publication.

In summary, it appears that many of the unusual properties of the A15 
materials are describable in terms of ``standard'' LDA electronic structure 
calculations when sublattice relaxation effects are included.  We have 
shown that the A15 materials (V$_3$Si, V$_3$Ge, and Nb$_3$Sn) are 
intrinsically unstable against shears 
producing the moduli $C_{11}-C_{12}$ and $C_{44}$, at $T=0$. Our calculations
also indicate that the zone center phonons ($\Gamma_2$ and $\Gamma_{12}$) are 
either unstable or extremely soft at $T=0$. Similar studies of the full 
phonon spectrum and the superconducting tunneling functions for the A15 
materials are being formulated to verify whether the ``standard'' LDA methods 
are consistent with all of the experimental data.

As a word of caution, we urge that elastic stability of a ``computer-designed''
novel material should always be checked, as one would not have anticipated the 
calculated instabilities in the A15 materials by examining the total energy 
verses volume curves alone.

The authors appreciate helpful discussions with M. J. Mehl, and gratefully 
acknowledge the support of the Campus Laboratory Collaboration Program of 
the University of California.

{\newpage
\widetext
\begin{table}
\caption{LAPW calculated and experimental elastic constants (in GPa)
(from Reference~\protect\onlinecite{test73}). Calculations were performed at 
the experimental volume using orthorhombic (for $C_{11}-C_{12}$)
and monoclinic (for $C_{44}$) distorted cells.  ``Unrelaxed'' refers to 
calculations in which atomic positions are held fixed at their cubic, ideal 
sites, while ``relaxed'' refers to calculations in which sublattices are 
allowed to relax. The elastic constants were extracted from a least square fit
of distortion energies to the form: $\Delta E=k_2\delta^2+k_4\delta^4$.
``Transforming'' (``Non Transforming'') denotes an experimental sample that
undergoes (does not undergo) a cubic to tetragonal phase transition.  The 
``unrelaxed'' $C_{11}-C_{12}$ values can also be extracted using a tetragonal 
distortion: such extracted values are within 6\% of those reported here.
}
\label{t-elastic}
\begin{tabular}{lccddcddcdd}
 & &\multicolumn{3}{c}{V$_3$Si}&\multicolumn{3}{c}{V$_3$Ge}
 &\multicolumn{3}{c}{Nb$_3$Sn}\\
 & Temp (K) & $B$ & $\frac{C_{11}-C_{12}}{2}$ & $C_{44}$ & $B$ & 
 $\frac{C_{11}-C_{12}}{2}$ & $C_{44}$ & $B$ & $\frac{C_{11}-C_{12}}{2}$ 
 & $C_{44}$ \\
\tableline
Calc. (Unrelaxed)     & 0   & 165 & 84.   & 79.   & 169 & 93. & 77.  & 159 & 91. & 53. \\
Expt. (Transforming)    & 300 & 176 & 83.8  & 81.0  &     &     & 69.8 & 154 & 68. & 40.8 \\
Expt. (Non Transforming) & 300 & 171 & 86.   & 80.9  & 170 & 92. & 70.3 & 160 & 70.7 & 39.6 \\
Calc. (Relaxed)       & 0   & 165 & $<$0. & $<$0. & 169 &$<$0.&$<$0. & 159 & $<$0. & $<$0. \\
Expt. (Transforming)    & 4.2 & 177 &  1.5  & 76.1  &     &     & 73.8 & 161 & 112. & 22.8 \\
Expt. (Non Transforming) & 4.2 & 171 &  8.   & 76.6  & 173 & 98.3& 72.3 & 165 & 0. & 26.6 \\
\end{tabular}
\end{table}
}

\end{document}